\documentclass[aps,prb,twocolumn,groupedaddress,floatfix,showpacs]{revtex4-2}
\usepackage[T1]{fontenc}
\usepackage{graphicx}
\usepackage{amsmath}
\usepackage{color}
\newcommand{\parl}{\parallel}
\begin{document}


\title{Electro-optical properties of excitons in Cu$_2$O
quantum wells: I discrete states}


\author{David Ziemkiewicz}
\email{david.ziemkiewicz@utp.edu.pl}
\author{Gerard Czajkowski}
\author{Karol Karpi\'{n}ski}
\author{Sylwia
Zieli\'{n}ska-Raczy\'{n}ska}
 \affiliation{Institute of
Mathematics and Physics, UTP University of Science and Technology,
\\Aleje Prof. S. Kaliskiego 7, 85-789 Bydgoszcz, Poland.}


\date{\today}

\definecolor{green}{rgb}{0,0.8,0}

\begin{abstract}
We present  theoretical results of the calculations of
optical functions for Cu$_2$O quantum well (QW) with Rydberg
excitons in an external homogeneous electric field of an arbitrary
field strength. Two configurations of an external electric field  
perpendicular and parallel to the QW planes  are considered in the energetic region for discrete excitonic states.
With the help of
the real density matrix approach,
which enables the derivation of the analytical expressions for the QW
electro-optical functions, absorption spectra are calculated  for the case of the excitation energy below  the  gap energy.

\end{abstract}
 \maketitle
\section{Introduction}

Excitons are of great physical interest since they represent the fundamental optical excitation in semiconductors. In particular, excitons in Cu$_2$O have attracted lots of attention \cite{AssmannBayer_2020}  in recent years due to an experiment, in which the hydrogenlike absorption
 spectrum of these quasiparticles up to a principal quantum number $n=25$ has been observed \cite{Kazimierczuk}. 
Since 2014 astonishing properties of these giant Rydberg excitons (RE) have been studied mostly in bulk systems in context of  their spectroscopic chracteristic as well as their linear and nonlinear interparticle interactions  and applications in quantum information technology \cite{Khazali}$
^-$\cite{my_kerr}. Most studies of
RE in an external electric field are concentrated
on the excitation energies below the fundamental gap in Cu$_2$O
\cite{SZR_2016,Heckotter_2017}. 

 The first experiments related to  properties of RE focused on natural
Cu$_2$O bulk crystals due to major difficulties in growing high-quality
synthetic samples. In the last few years the technological
progress enabled the growth of Cu$_2$O microcrystals with excellent
optical material quality and very low point defect levels \cite{Stainhauer}. This enabled Cu$_2$O based low-dimensional systems (quantum wells, wires, and dots) to be realized experimentally  \cite{Naka}$^-$\cite{Hamid}.

 Cuprous oxide
is a semiconductor characterized by large exciton binding energy and with
significant technological importance in applications such as
photovoltaics and solar water splitting. It is also a superior
material system for quantum optics that might enable  observation of
 Rydberg excitons in nano structures. 
Motivated by technological development and potential applications, investigations of RE in low dimensional systems  have started recently \cite{Konzelmann, Ziemkiewicz_2020, Ziemkiewicz_2021}.
In our previous papers  with the help od real density matrix approach (RDMA) we have considered optical properties of RE in quantum dots and quantum wells (QW)\cite{Ziemkiewicz_2020} and later we have studied Rydberg magnetoexcitons in QW \cite{Ziemkiewicz_2021}. The applied approach turned out to be useful to describe the fine structure splitting of excitons lines in absorption spectra for any magnetic field strength.
The natural step forward at this moment is to study optical response of RE in quantum wells subjected to an  interaction with the electric field.for the excitation energy below the gap. In such a situation one can distinguish two cases regarding directions of this external electric field, which can be oriented parallel or perpendicular to the quantum well layer. The first case resembles that known from the bulk  in an electric field of the energy below the gap \cite{SZR_2016}; the degenerations of excitonic levels are lifted, increasing number of peaks corresponding to increasing state number appears, resonances are shifted and anticrossings of lines are observed. For the electric field perpendicular to the quantum well layer the situation is quite distinct from that in bulk semiconductor. The electron and the hole creating the exciton are attracted by their Coulomb attraction and they are confined in a plane of the quantum well and as a consequence large Stark shifts of excitons absorption peaks appear; this phenomenon is called Quantum-Confined Stark Effect; for recent review see \cite{kuo, GC}. Bellow we will considered these two cases in details.

The paper is organized as follows. In Sec. II we recall the basic equations of the RDMA, adapted for the case of QWs when an external electric field is applied. The Section II is divided into two parts, where different configurations are considered. In Subs. A we consider vthe case of the electric field applied parallel to the z axis (the growth axis), i.e. perpendicular to the QW planes, and in Subs. B the case of the lateral electric field. In both cases we derive analytical expressions for the  QW mean effective electro-susceptibility. Those expressions are then used in Sect. III where detailed calculations for the Cu$_2$O based QWs are presented. The summary and conclusions of our paper are presented in Sec. IV. The Appendices  A,B contain the details of the analytical calculations.

\section{Theory}
We consider a Cu$_2$O quantum well  of thickness $L$, located
in the $xy$ plane, with QW surfaces located at $z=\pm L/2$. A linearly
polarized electromagnetic wave of the frequency $\omega$ is incident normally on the
QW. The wave vector has only one component $\bf{k} = k_z$ and the electric field vector ${\bf E}=E_x$.

We aim to discuss the changes of the QW optical response when
a constant external electric field \textbf{F} is applied.  The polarization of
electrons and holes induced by this field leads to a significant
decrease of the exciton binding energy.
 As it was pointed out there are two opposite  directions
  in which one can applied an electric field to QW: an external field is parallel to the  the layers
  or with the  field is directed perpendicular to the layer. In  the following subsections both cases will be
  discussed. As in the previous papers \cite{Ziemkiewicz_2020, Ziemkiewicz_2021}, we use the real density matrix approach  for
  calculating the QW optical functions (absorption, reflection, and
  transmission). In particular, the RDMA turned out to be  appropriate for
  computing the effects of external fields since it includes both
  the relative motion of the carriers and the center-of-mass
  motion, where the interaction with the radiation takes place. 
This approach allows also for including the band mixing effects originate from lifting  degenerations of states caused by an external electric field.

\subsection{The electric field  parallel to the $z$-axis} We use
the RDMA approach, as described in ref. \cite{Ziemkiewicz_2021} to
determine the electro-optical properties. The starting point is
the constitutive equation
\begin{equation}\label{constitutiveHL}
(H_{QW}-\hbar\omega-i{\mit\Gamma})Y=\textbf{ME},
\end{equation}
with the two-band QW Hamiltonian
\begin{eqnarray}\label{BFmagnetichamiltonian1}
&&\phantom{nucl}H_{QW}=E_{g}+\frac{1}{2m_e} \left({\bf p}_e-e
\frac{{\bf r}_e \times {\bf B}}{2}\right)^2 \nonumber\\&&+
\frac{1}{2m_{h}} \left({\bf p}_h + e \frac{{\bf r}_h \times {\bf
B}}{2}\right)_z^2 \nonumber\\ & &+ \frac{1}{2m_{h}} \biggl( {\bf
p}_h+e \frac{{\bf r}_h \times {\bf B}}{2}\biggr)_\parl^2\\&&
+e{\bf F}\cdot({\bf r}_e-{\bf r}_h) + V_{\rm conf}({\bf r}_e,{\bf
r}_h) - \frac{e^2}{4\pi\epsilon_0\epsilon_b \vert{\bf r}_e - {\bf
r}_h\vert} ,\nonumber
\end{eqnarray}
 \noindent
 {\bf B} is the magnetic field
vector, ${\bf F}$ the electric field vector,  $V_{\rm conf}$ are
the surface potentials for electrons and holes, $m_{h}, m_{e}$ are
the hole and the electron effective masses. We separate
 the exciton center-of-mass and relative motion,
 and consider the case of ${\bf B}=0$, for $\textbf{F}\parallel
 z$ and the dipole density ${\bf M}\parallel{\bf E}$.
The electron-hole interaction is used in the two-dimensional
approximation which enables to obtain the  solutions
in an analytical form. The calculations of electro-optical properties
become much simpler when we consider a quantum well of parabolic
confinement potentials in the form of an harmonic oscillator
potential
$V_{\rm
conf}=\frac{1}{2}m_{e}\omega_{ez}^2z_e^2+\frac{1}{2}m_{h}\omega_{hz}^2z_h^2$,
\noindent where the energies $\hbar\omega_{ez}, \hbar\omega_{hz}$
correspond to the electron and hole barriers. For the considered geometry  the QW Hamiltonian has
the form
\begin{eqnarray}\label{QWparabHamilt1}
&&\phantom{nuc}H_{\rm
QW}=E_g+H_{m_{e},\omega_{ez}}^{(1D)}(z_e)+H_{m_{h},\omega_{hz}}^{(1D)}(z_h)\nonumber\\&&+
H_{\rm Coul}^{(2D)}(\hbox{\boldmath$\rho$})+eF(z_e-z_h),\nonumber
\end{eqnarray}
\noindent and  contains the one-dimensional oscillator Hamiltonians
\begin{equation}\label{1dimoschamiltonian1}
H^{(1D)}_{m,\omega}(z)=\frac{p_z^2}{2m}+\frac{1}{2}m\omega^2z^2,
\end{equation}
\noindent and the two-dimensional Coulomb Hamiltonian
\begin{equation}\label{2dimCoulombhamilt1}
H_{\rm Coul}^{(2D)}(\hbox{\boldmath$\rho$})=\frac{{\bf
p}^2_{\parallel}}{2\mu_{\parallel}}-\frac{e^2}{4\pi\epsilon_0\epsilon_b\rho}.
\end{equation}

\noindent Using the substitution
\begin{eqnarray}\label{zetaezetah}
&&\zeta_{e}=z_e+z_{0e},\quad z_{0e}=\frac{eF}{m_{e}\omega_{ez}^2},\nonumber\\
&&\zeta_{h}=z_h-z_{0h},\quad z_{0h}=\frac{eF}{m_{h}\omega_{hz}^2},
\end{eqnarray}
\noindent we obtain that the QW Hamiltonian (\ref{QWparabHamilt1}) can be rewritten as
\begin{eqnarray}\label{QWparabHamiltdisp1}
&&\phantom{nuc}H_{\rm
QW}=E_g+H_{m_{ez},\omega_{ez}}^{(1D)}(\zeta_e)+H_{m_{h},\omega_{hz}}^{(1D)}(\zeta_h)\nonumber\\
&&+ H_{\rm
Coul}^{(2D)}(\hbox{\boldmath$\rho$})-\frac{(eF)^2}{2m_{e}\omega_{ez}^2}
-\frac{(eF)^2}{2m_{h}\omega_{hz}^2}.
\end{eqnarray}
\noindent 
 Using the long wave approximation we  seek  solutions of Eq.(1) in the
form
\begin{eqnarray}\label{rozwY}
&&Y(\rho,\zeta_e,\zeta_h)=\\
&&=E(Z)\sum_{jm N_eN_h}
c_{jmN_eN_h}\psi_{jm}(\hbox{\boldmath$\rho$})
\psi^{(1D)}_{\alpha_{e},N_e}(\zeta_e)\psi^{(1D)}_{\alpha_{h},N_h}(\zeta_h),\nonumber
\end{eqnarray}
\noindent where $\psi_{jm}$ are the normalized eigenfunctions of
the 2-dimensional Coulomb Hamiltonian,
\begin{eqnarray}\label{2_Dim_Eigen}
 &&\psi_{jm}(\rho,\phi)=R_{jm}(\rho)\frac{e^{im\phi}}{\sqrt{2\pi}},\nonumber\\
&&R_{jm}=A_{jm}e^{-2\lambda\rho}(4\lambda\rho)^{\vert
m\vert}L^{\vert 2
m\vert}_j(4\lambda\rho),\\
&&\lambda=\frac{1}{1+2(j+\vert m\vert)},\nonumber\\
&&A_{jm}=\frac{4}{(2j+2\vert
m\vert+1)^{3/2}}\left[\frac{j!}{(j+2\vert
m\vert)!}\right]^{1/2},\nonumber
\end{eqnarray}
and $L^\alpha_n(x)$ are the Laguerre polynomials, for which we
use the definition
\begin{displaymath}
L^\alpha_n(x)={n+\alpha\choose
n}M(-n,\alpha+1;x),\end{displaymath} with  the Kummer function
$M(a,b,z)$ (the confluent hypergeometric
function)\cite{Abramowitz}, $\rho=r/a^*$ is the scaled space
variable.
 and
$\psi^{(1D)}_{\alpha,N}(z)$ (N=0,1,...) are the quantum oscillator
eigenfunctions of the Hamiltonian (\ref{1dimoschamiltonian1})
\begin{eqnarray*}\label{eigenf1doscillator}
&&
 \psi^{(1D)}_{\alpha,N}(z)=
 \pi^{-1/4}\sqrt{\frac{\alpha_{z}}{2^N N!}} H_N(\alpha z)
e^{-\frac{\alpha^2}{2}z^2}, \\
&&
 \alpha = \sqrt{\frac{m \omega_{z}}{\hbar}},
\end{eqnarray*}
 with  Hermite polynomials $H_N(x)$, $(N=0,1,\ldots)$.

\noindent Here we use the transition dipole density in the form
\cite{Ziemkiewicz_2021}
\begin{equation}\label{dipoledensity}
M(\hbox{\boldmath$\rho$},z_e,z_h)=\frac{M_0}{2\rho_0^3}\rho\,e^{-\rho/\rho_0}\frac{e^{i\phi}}{\sqrt{2\pi}}\delta(z_e-z_h).
\end{equation}
$M_0$ is the integrated strength, the coherence radius is defined
$\rho_0=r_0/a^*$, with $r_0=\sqrt{\frac{\hbar^2}{2\mu E_g}}$ and $a^*$ is the excitonic Bohr radius. These coefficients 
 are connected
through the longitudinal-transversal energy $\Delta_{LT}$ 
\begin{equation}
(M_0\rho_0)^2=\frac{4}{3}\frac{\hbar^2}{2\mu}\epsilon_0\epsilon_ba^*\frac{\Delta_{LT}}{R^*}\,e^{-4\rho_0}.
\end{equation}
Assuming that the electromagnetic wave of the  component
$E(Z)$ is linearly polarized, we substitute $Y$ from Eq. (\ref{rozwY}) into Eq.
(\ref{constitutiveHL}) to calculate the expansion coefficients
$c_{jmN_eN_h}$
\begin{eqnarray}
&&\sum_{jm N_eN_h}
c_{jmN_eN_h}\biggl[E_g-\hbar\omega-i{\mit\Gamma}\nonumber\\
&&+\left(N_e+\frac{1}{2}\right)\hbar\omega_{ez}+
\left(N_e+\frac{1}{2}\right)\hbar\omega_{ez}\\
&&-\frac{(eF)^2}{2m_{ez}\omega_{ez}^2}
-\frac{(eF)^2}{2m_{hz}\omega_{hz}^2}\biggr]
\psi_{jm}(\hbox{\boldmath$\rho$})
\psi^{(1D)}_{\alpha_{e},N_e}(\zeta_e)\psi^{(1D)}_{\alpha_{h},N_h}(\zeta_h)\nonumber\\
&&=M(\hbox{\boldmath$\rho$},z_e,z_h),\nonumber
\end{eqnarray}
obtaining
\begin{eqnarray*}
&&c_{j1N_eN_h}=\langle \Psi_{N_eN_h}\rangle_\infty
b_{j1}\biggl[E_g-\hbar\omega-i{\mit\Gamma}+W_{N_e}+W_{N_h}\\ &&
-\frac{(eF)^2}{2m_{ez}\omega_{ez}^2}
-\frac{(eF)^2}{2m_{hz}\omega_{hz}^2}\biggr]^{-1},\end{eqnarray*}
with the following definitions 
\begin{eqnarray}\label{definitions}
&&\langle \Psi_{N_eN_h}\rangle_\infty=\int_{-\infty}^\infty
\psi^{(1D)}_{\alpha_{e},N_e}(\zeta_e)\psi^{(1D)}_{\alpha_{h},N_h}(\zeta_h)dz\\
 &&=\sqrt{\frac{\alpha_e\alpha_h}{\pi\,2^{N_e+N_h}N_e!\;N_h!}}\nonumber\\
 && \times\int\limits_{-\infty}^\infty
 dz\biggl\{\,H_{N_e}[\alpha_e(z+z_{0e})]e^{-\frac{\alpha_e^2(z+z_{0e})^2}{2}}\nonumber\\
 &&\times\,H_{N_h}[\alpha_h(z-z_{0h})]e^{-\frac{\alpha_h^2(z-z_{0h})^2}{2}}\biggr\},\nonumber\\
 &&W_{N_e}=\left(N_e+\frac{1}{2}\right)\hbar\omega_{ez},\nonumber\\
 &&W_{N_h}=\left(N_h+\frac{1}{2}\right)\hbar\omega_{hz},\nonumber\\
 &&b_{j1}=\frac{2\mu}{\hbar^2}\left\{(M_0\rho_0)\frac{6}{\sqrt{2}}\right.\\
&&\left.\times\sqrt{\frac{(j+1)(j+2)}{(j+3/2)^5}}(1+2\rho_0\lambda_{j1})^{-4}
F\left(-j,4;3;\frac{1}{s}\right)\right\}\nonumber\\
&&s=\frac{1+2\rho_0\lambda_{j1}}{4\rho_0\lambda_{j1}},\nonumber\\
&&\lambda_{j1}=\frac{1}{2j+3},\nonumber
 \end{eqnarray}
 and $F(\alpha,\beta;\gamma;z)$ denotes a hypergeometric
 series\cite{Abramowitz} (in  Ref.\cite{Ziemkiewicz_2021} 
 the calculation of $b_{j1}$ is elaborately presented).  In the RDMA,
the total polarization of the medium is related to the coherent
amplitude $Y$ by
\begin{equation}\label{Polar}
{\bf P}({\bf R})=2 \hbox{Re}\int d^3{r}\,{\bf M}({\bf r}) Y({\bf
R},{\bf r})
\end{equation}
where $\textbf{R}$ is the center-of-mass coordinate. This, in
turn, is used in the Maxwell's  equation
\begin{equation}\label{Maxwell}
c^2\nabla^2 {\bf E(R)} - \epsilon_b \ddot{\bf E} =
\frac{1}{\epsilon_0}{\bf \ddot{P}(R)}.
\end{equation}
Using the long wave approximation we obtain the coherent amplitude
$Y$ from Eq. (\ref{constitutiveHL}) as linearly dependent on the
electric field \textbf{E}. Then, from Eq. (\ref{Polar}), one can
determine the  susceptibility
$\chi(\textbf{R})$\cite{Ziemkiewicz_2021}. 

\noindent For linearly polarized
wave, in the considered configuration for the wave propagating in
the $z$-direction, we consider one component $E(Z)$ and one $P(Z)$
of the electric and polarization vectors, obtaining the position-dependent
susceptibility 
$\chi(Z)=\frac{P(Z)}{\epsilon_0E(Z)}$.
  \,\,\,Below we will use
the mean effective QW susceptibility
\begin{equation}\label{meaneffchi}
\chi=\frac{1}{L}\int\limits_{-L/2}^{L/2}\frac{P(Z)}{\epsilon_0E(Z)}\;dz.\end{equation}
For the  considered case of the electric field perpendicular to the QW layer, the polarization determined from Eq.
(\ref{Polar}), regarding also the form of $\textbf{M}$, has the
form
\begin{eqnarray}\label{QW_Polar}
&&P(Z)=2M_0\sum\limits_{N_e=0}^{N_{\hbox{\tiny{\emph{e}
max}}}}\sum\limits_{N_h=0}^{N_{\hbox{\tiny{\emph{h}
max}}}}\sum\limits_{j=0}^J\biggl\{c_{j1N_eN_h}b_{j1}\nonumber\\
&&\times\psi^{(1D)}_{\alpha_{e},N_e}(Z+z_{e0})
\psi^{(1D)}_{\alpha_{h},N_h}(Z-z_{h0})\biggr\}.\nonumber
\end{eqnarray}
Here $J$ denotes the upper limit of $j$, which corresponds to the number of excitonic states taken into account. Using  Eq. (\ref{meaneffchi}) we arrive at the equation
determining the mean effective susceptibility for a QW, when the
homogeneous electric field F is applied perpendicular to the QW
plane
\begin{eqnarray}\label{chi2D}
&&\chi^{(2D)}(\omega)=48\epsilon_b\frac{\Delta_{LT}}{R^*}\left(\frac{a^*}{L}\right)
\,\sum\limits_{N_e=0}^{N_{\hbox{\tiny{\emph{e}
max}}}}\sum\limits_{N_h=0}^{N_{\hbox{\tiny{\emph{h}
max}}}}\sum\limits_{j=0}^J \\
&&\times\frac{
f_{j}^{(2D)}\langle\Psi_{N_eN_h}\rangle_\infty\,\langle\Psi_{N_eN_h}\rangle_L}{L(E_g-\hbar\omega+E_{j1}+W_{N_e}+W_{N_h}+\Delta
E- i{\mit\Gamma}_{jN_eN_h})},\nonumber
\end{eqnarray}
where 
\begin{eqnarray}\label{definitions_chi}
&&f_{j1}^{(2D)}={48}\frac{(j+1)(j+2)}{\left(j+\frac{3}{2}\right)^5}\frac{\left[F\left(-j,4;3;\frac{4\lambda_{j1}\rho_0}{1+2\lambda_{j1}\rho_0}\right)\right]^2}{(1+2\lambda_{j1}\rho_0)^8}
,\nonumber\\
&&E_{jm}=-\frac{4}{(2j+2\vert m\vert +1)^2}R^*,\\
&&\langle\Psi_{N_eN_h}\rangle_L=\nonumber\\
 &&=\sqrt{\frac{\alpha_e\alpha_h}{\pi\,2^{N_e}N_e!\;2^{N_h}N_h!}}\int_{-L/2}^{L/2}
 dz\,\biggl\{H_{N_e}(z+z_{0e})\nonumber\\
 &&\times
 e^{-\alpha_e^2(z+z_{0e})^2/2}\,H_{N_h}(z-z_{0h})e^{-\alpha_h^2(z-z_{0h})^2/2}\biggr\},\nonumber
 \end{eqnarray}
The Stark shift is given by
 \begin{equation}\label{Stark_shift}
\Delta
E=-\frac{e^2F^2}{2m_{ez}\omega_{ez}^2}-\frac{e^2F^2}{2m_{hz}\omega_{hz}^2}.
\end{equation}
 For further calculations we have to define the confinement
 parameters $\alpha_{e,\alpha_h}$. 
 We identify the oscillator
 energies $W_{N_e=0}, W_{N_h=0}$ with those of the lowest energies
 of the infinite well potentials
 \begin{equation}
 W_{N_e=0}=\frac{\hbar^2}{2m_e}\frac{\pi^2}{L^2},\quad
 W_{N_h=0}=\frac{\hbar^2}{2m_h}\frac{\pi^2}{L^2},
 \end{equation}
 which gives the coefficients
 \begin{equation}\label{alpha}
 \alpha_{e}=\alpha_{h}=\alpha=\frac{1}{a^*}\left(\frac{\pi}{L}\right).\end{equation}
With so chosen confinement parameters the  explicit expressions for $\langle
\Psi_{N_eN_h}\rangle_\infty$ and $\langle \Psi_{N_eN_h}\rangle_L$
for the lowest combinations of the quantum numbers $N_e,N_h$ are derived in Appendix
\ref{Appendix A}. The
Stark shift (\ref{Stark_shift}) expressed by the confinement
parameters depends also on the total excitonic mass and the
applied field strength $f=\frac{F}{F_{\rm
 I}}$ ($F_I=\frac{R^*}{ea^*}$   is the ionization field)
\begin{equation}
\Delta E=-\frac{1}{4\pi^4}f^2\left(\frac{M_{\rm
 tot}}{\mu}\right)\left(\frac{L}{a^*}\right)^4\,R^*.
\end{equation}
For Cu$_2$O the ionization field is quite large (due to the smallness of the excitonic Bohr radius), so a feasible cases always correspond to $f<<1$. Such range of the field strength is discussed in our paper.
As follows from Eq. (\ref{chi2D}),  the applied
 electric field in this configuration causes the appearance of
 confinement states with $N_e\neq N_h$, which, in the model with
 equal confinement parameters for electron and hole, are absent in
 the case without field. As illustration, we present the formula
 for the mean effective electro-susceptibility, where the lowest
 confinement states $(N_e=N_h=0), (N_e=1,N_h=0), (N_e=0,N_h=1)$,
 and $J$ 2D exciton states are accounted for
 \begin{eqnarray*}\label{chi2D_Z}
&&\chi^{(2D)}(\omega)\nonumber\\
&&=\sum\limits_{j=0}^J \frac{\epsilon_b\Delta_{LT}a^*
f_{j}^{(2D)}\langle\Psi_{00}\rangle_\infty\,\langle\Psi_{00}\rangle_L}{L(E_g-\hbar\omega+E_{j1}+W_{N_e=0}+W_{N_h=0}+\Delta
E-
i{\mit\Gamma}_{j00})}\nonumber\\
&&+\sum\limits_{j=0}^J \frac{\epsilon_b\Delta_{LT}a^*
f_{j}^{(2D)}\langle\Psi_{10}\rangle_\infty\,\langle\Psi_{10}\rangle_L}{L(E_g-\hbar\omega+E_{j1}+W_{N_e=1}+W_{N_h=0}+\Delta
E-
i{\mit\Gamma}_{j10})}\nonumber\\
&&+\sum\limits_{j=0}^J \frac{\epsilon_b\Delta_{LT}a^*
f_{j}^{(2D)}\langle\Psi_{01}\rangle_\infty\,\langle\Psi_{01}\rangle_L}{L(E_g-\hbar\omega+E_{j1}+W_{N_e=0}+W_{N_h=1}+\Delta
E-
i{\mit\Gamma}_{j01})},\nonumber\\
\end{eqnarray*}
The expressions for the quantities
$\langle\Psi_{00}\rangle_\infty\,\langle\Psi_{00}\rangle_L,\,\langle\Psi_{10}\rangle_\infty\,\langle\Psi_{10}\rangle_L$,
and $\langle\Psi_{01}\rangle_\infty\,\langle\Psi_{01}\rangle_L$
are given in the Appendix \ref{Appendix A}. All relevant parameters are summarized in the Table \ref{tab1}; The dissipation constant $\Gamma$ decreases with $j$ according to experimental data in Ref.\cite{Kazimierczuk}.
\begin{table}[ht!]
\caption{\small Band parameter values for Cu$_2$O, Rydberg energy
and excitonic radius calculated from effective masses; masses in free electron
mass $m_0$, the ionization field $F_{\rm I}=R^*/(ea^*)$}\label{tab1}
\begin{center}
\begin{tabular}{p{.2\linewidth} p{.2\linewidth} p{.2\linewidth} p{.2\linewidth} p{.2\linewidth}}
\hline
Parameter & Value &Unit&Reference\\
\hline $E_g$ & 2172.08& meV&\cite{Ziemkiewicz_2020}\\
$R^*$&87.78&meV&\cite{Ziemkiewicz_2020}\\
$\Delta_{LT}$&$1.25\times 10^{-3}$&{meV}&\cite{Ziemkiewicz_2020}\\
$\mit\Gamma$&$3.88/(j+1)^3$&meV&\cite{maser}\\
$m_e$ & 1.0& $m_0$&\cite{Ziemkiewicz_2020}\\
$m_h$ &0.7&  $m_0$&\cite{Ziemkiewicz_2020}\\
$M_{\hbox{\tiny tot}}$&1.56&$m_0$&\cite{Ziemkiewicz_2020}\\
$\mu$ & 0.363 &$m_0$&\cite{Ziemkiewicz_2020}\\
$a^*$&1.1& nm&\cite{Ziemkiewicz_2020}\\
$r_0$&0.22& nm&\cite{Ziemkiewicz_2020}\\
$\epsilon_b$&7.5 &&\cite{Ziemkiewicz_2020}\\
 $F_{\rm I}$&1.02$\times\,10^{3}$&kV/cm\\
\hline
\end{tabular} \label{parametervalues1}\end{center}
\end{table}
\subsection{The electric field  parallel to x axis}
In this section we will discuss the case of the electric field \textbf{F} applied parallel to the layer and still we will consider the case of 
 the excitation energy
$\hbar\omega$ smaller than the band gap. The electric field can be
considered as a perturbation  and methods  similar to that
used in our previous papers  for the electric
field applied to bulk crystal \cite{SZR_2016} or  for
the magnetoexcitons in a QW  \cite{Ziemkiewicz_2021} can be used. Assuming  the same shape of the
confinement parabolic e-h potential 
and the
two-dimensional Coulomb interaction the QW
Hamiltonian consists of the following operators

\begin{eqnarray}\label{QWparabHamilt2}
&&H_{\rm
QW}=E_g+H_{m_{e},\omega_{ez}}^{(1D)}(z_e)+H_{m_{h},\omega_{hz}}^{(1D)}(z_h)\nonumber\\
&&+ H_{\rm Coul}^{(2D)}(\hbox{\boldmath$\rho$})+eF(x_e-x_h).
\end{eqnarray}
Considering the term $eF(x_e-x_h)$ as a perturbation, we seek the
solution of the constitutive equation in terms of the
eigenfunctions of the unperturbed part of the Hamiltonian
\begin{eqnarray}\label{Yparal}
&&Y=\\
&&=\sum\limits_{jmN_eN_h}c_{jmN_eN_h}\psi^{(2D)}_{jm}(\rho,\phi)\psi^{(1D)}_{\alpha_e,N_e}(z_e)\psi^{(1D)}_{\alpha_h,N_h}(z_h).\nonumber
\end{eqnarray}
The functions are defined in the previous section. Proceeding in similar way as it was done in Subsection A, i.e. substituting
the expansion (\ref{Yparal}) into the constitutive equation
(\ref{constitutiveHL}) with the Hamiltonian
(\ref{QWparabHamilt2}) and the dipole density
(\ref{dipoledensity}), we arrive to a set of equations for the
expansion coefficients 
\begin{eqnarray}\label{eqslateral}
&&\sum\limits_{j=0}^{J-1}\sum\limits_{N_eN_h}c_{j1N_eN_h}\kappa^2_{j1N_eN_h}\delta_{ij}\delta_{N_eN_h}
\nonumber\\&&+\sum\limits_{j=0}^{J-1}\sum\limits_{N_eN_h}c_{j 0
N_eN_h}V^{10}_{ij
}\delta_{N_eN_h} \nonumber\\
&&+\sum\limits_{j=0}^{J-1}\sum\limits_{N_eN_h}c_{j 2
N_eN_h}V^{12}_{ij}\delta_{N_eN_h}\nonumber\\&&
=\frac{2\mu}{\hbar^2}\frac{1}{a^*}\langle R_{i1}\vert M\rangle\langle\Psi_{N_eN_h}\rangle_\infty,\nonumber\\
&&i=0,1,\ldots, J-1,\nonumber\\
\end{eqnarray}
and
\begin{eqnarray}
&&\sum\limits_{j=0}^{J-1}\sum\limits_{N_eN_h}c_{j0N_eN_h}\kappa^2_{j0N_eN_h}\delta_{ij}\delta_{N_eN_h}\nonumber\\
&&+2\sum\limits_{j=0}^{J-1}\sum\limits_{N_eN_h}\,c_{j 1 N_eN_h}\,
V^{01}_{ij}\delta_{N_eN_h}=b_i;\nonumber\\ &&i=J,J+1,\ldots
\nonumber\\
&&\sum\limits_{j=0}^{J-1}\sum\limits_{N_eN_h}c_{j2N_eN_h}\kappa^2_{j2N_eN_h}\delta_{ij}\delta_{N_eN_h}\nonumber\\
&&+2\sum\limits_{j=0}^{J-1}\sum\limits_{N_eN_h}\,c_{j 1 N_eN_h}\,
V^{21}_{ij}\delta_{N_eN_h}=b_i;\nonumber\\&& i=2J,2J+1,\ldots\nonumber
\end{eqnarray}
where the following definitions were used
\begin{equation}
\kappa^2_{jmN_eN_h^2}=\frac{1}{R^*}\left(E_g-\hbar\omega+E_{jm}+W_{N_e}+W_{N_h}-i{\mit\Gamma}\right).
\end{equation}
 $V$ are the matrix elements
\begin{eqnarray}
&&V_{\ell s}^{01}=\frac{1}{2}f\int\limits_0^\infty
\rho^2\,d\rho\,R_{\ell 0}(\rho)R_{s1}(\rho),\nonumber\\
&&V_{\ell s}^{10}=\frac{1}{2}f\int\limits_0^\infty
\rho^2\,d\rho\,R_{\ell 1}(\rho)R_{s0}(\rho),\\
&&\ell, s=0,1,\ldots,J-1,\nonumber\\
&& V_{\ell s}^{12}=\frac{1}{2}f\int\limits_0^\infty
\rho^2\,d\rho\,R_{\ell 1}(\rho)R_{s2}(\rho).\nonumber
\end{eqnarray}
The equations (\ref{eqslateral}) form the set of 3$J$ linear algebraic
equations. They can be put into a matrix form
\begin{eqnarray}\label{matrixform}
&&\underline{\underline{A}}\textbf{X}=\textbf{b},\nonumber\\
&&\textbf{X}=(x_1,x_2,\cdots x_{3J}),\\
&&\textbf{b}=(b_1,b_2,\cdots b_{3J}),\nonumber \end{eqnarray}
where the matrix elements $\underline{\underline{A}}$ and the
components of the vector \textbf{b} are defined in Appendix
\ref{Appendix_B}. With the solutions \textbf{X} one can obtain the
expression for the effective QW electro-susceptibility for
$F\parallel x$
\begin{eqnarray}\label{chix}
&&\chi^{(2D)}(\omega)\\
&&=48\epsilon_b\frac{\Delta_{LT}}{R^*}\left(\frac{a^*}{L}\right)\hbox{erf}\left(\frac{L\sqrt{p}}{2}\right)\frac{\alpha_e\alpha_h}{p}
\sum\limits_{i=1}^Jb_i\,x_i.\nonumber
\end{eqnarray}
In order to illustrate such approach and to have an extensive insight into this dependence we take the simplest case,
when $J=1.$ Here  Eq. (\ref{matrixform}) takes the form
\begin{eqnarray*}
\left(\begin{array}{ccc}
\kappa_{01}^2&V^{10}_{00}&V^{12}_{00}\\
2V_{00}^{01}&\kappa_{00}^2&0\\
2V_{00}^{21}&0&\kappa_{02}^2\\
\end{array}\right)
\left(
\begin{array}{c}
x_1\\x_2\\ x_{3}\end{array}\right) =\left(
\begin{array}{c}
b_1\\0 \\0\\
\end{array}\right),
\end{eqnarray*}
where
\begin{eqnarray*}
&&\kappa_{00}^2=\frac{1}{R^*}\left(E_g-\hbar\omega-i{\mit\Gamma}_0-4R^*+W_{e0}+W_{h0}\right)\\
&&=\frac{1}{R^*}\left(E_{T0}-E-i{\mit\Gamma}_0\right),\\
&&\kappa_{01}^2=\frac{1}{R^*}\left(E_g-\hbar\omega-i{\mit\Gamma}_1-\frac{4}{9}R^*+W_{e0}+W_{h0}\right)=\\
&&=\frac{1}{R^*}\left(E_{T1}-E-i{\mit\Gamma}_1\right),\\
&&\kappa_{02}^2=\frac{1}{R^*}\left(E_g-\hbar\omega-i{\mit\Gamma}_2-\frac{4}{25}R^*+W_{e0}+W_{h0}\right)=\\
&&=\frac{1}{R^*}\left(E_{T2}-E-i{\mit\Gamma}_2\right).\\
\end{eqnarray*}
The relevant root is given by
\begin{eqnarray}
&&x_1=\frac{b_1\kappa_{00}^2\kappa_{02}^2}{\Delta}, \end{eqnarray}
where \begin{eqnarray*} &&\Delta
=-E^3+E^2(E_{T0}+E_{T1}+E_{T2})-E(E_{T1}E_{T0}\\
&&+E_{T2}E_{T0}+E_{T1}E_{T2})
+E_{T0}E_{T1}E_{T2}-af^2,\\
&&E_{Tm}=\frac{1}{R^*}(E_g+E_{jm}),\\
&&E_{jm}=-\frac{4R^*}{(2j+2\vert m\vert +1)^2},\qquad j=0,\;m=0,1,2,\\
&&a=2\frac{V^{12}_{00}V_{00}^{21}+V_{00}^{01}V^{10}_{00}}{f^2}.
\end{eqnarray*}
Above expressions are used to determine the susceptibility (\ref{chix}), which
 shows resonant behaviour at the energies
resulting from the equation
\begin{equation}\label{eq:Dzero}
\hbox{Re}\,\Delta=0,
\end{equation}
which is a 3-rd degree equation for  $E=\hbar\omega/R^*$. The
constant term contains  $f^2$,  so the solutions will also
depend on $f^2$. Assuming that $f$ is small we can seek solutions
$X_a,X_b,X_c$ near the unperturbed values $E_{T0},E_{T1},E_{T2}$,
$X=X_a+\delta$, etc. Solving the equation (\ref{eq:Dzero}) and
retaining terms linear in $\delta$ one obtains
\begin{eqnarray}
&&\delta_a=\frac{af^2}{X_a(X_b+X_c-X_a)-X_bX_c},
\end{eqnarray}
and similarly
\begin{eqnarray}
&&\delta_b=\frac{af^2}{X_b(X_a+X_c-X_b)-X_aX_c},\nonumber\\
&&\delta_c=\frac{af^2}{X_c(X_a+X_b-X_c)-X_aX_b}.
\end{eqnarray}
The above quantities  correspond to $m=0,1,2$, which characterize $s$, $p$ and $d$ excitons. Despite the nontrivial form of Eq. (\ref{chix}) and its dependence on field $f$, some general conclusions can be formulated: the energy shift $\delta_{a,b,c}$ is proportional to $f^2$, with quantity $\delta$ varies from state to state, whereas in the
configuration  $\textbf{F}\parallel z$ the Stark shit has a constant sign and depends only on the QW thickness. By substituting the values for Cu$_2$O, one obtains $X_a \approx 21.28$,~$X_b \approx 17.72,~X_c=21.56$ and correspondingly $\delta_a \approx 1$, $\delta_b \approx -13.67$, $\delta_c \approx -1.08$. Therefore, the shifts differ by signs (2 negative and 1 positive) and the most significant occurs for $p$ exciton. The anticrossings may originate from the opposite signs of shifts $\delta_{a,b,c}$; When the field
strength $f$ increases, for a certain critical value of $f$ the inversion of places appears: the level $p$ will be (in energetic scale) above the level $d$.

\section{Results and discussion}
\subsection{The \textbf{F} field perpendicular to the QW layer ($\textbf{F}||z$)}
The behaviour of electro-absorption for electric field perpendicular to quantum well layers is quite distinct from that in bulk semiconductors which is a straightforward consequence of the quantum well gist. Vividly speaking an electric field applied perpendicular to the layer pull the hole and electron (forming the exciton) in opposite directions squashing them against walls of quantum well and both particles are strongly attracted by their Coulomb interaction. Taking into account higher excitonic states one can observe that  the exciton absorption peaks are  broadened  and as a consequence of constrains due to QW  there appear large Stark shifts (towards lower energies), which is known as the Quantum-Confined Stark Effect.
\begin{figure}[ht!]
\centering
\includegraphics[width=.9\linewidth]{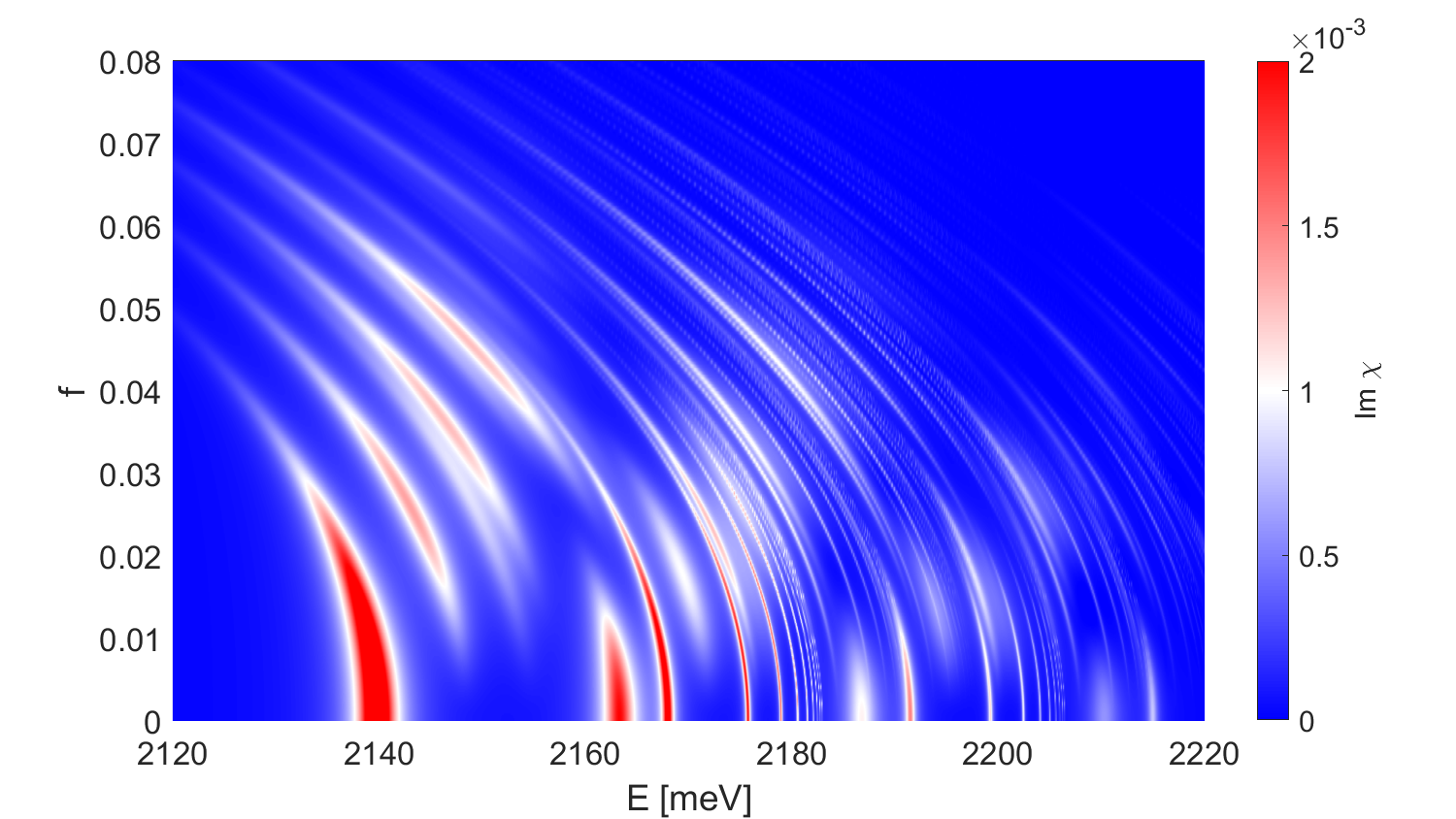}
\caption{Imaginary part of susceptibility as a function of energy and electric field $f$ for $L$=10 nm.}\label{Fig:1}
\end{figure}

\begin{figure}[ht!]
\centering
\includegraphics[width=1\linewidth]{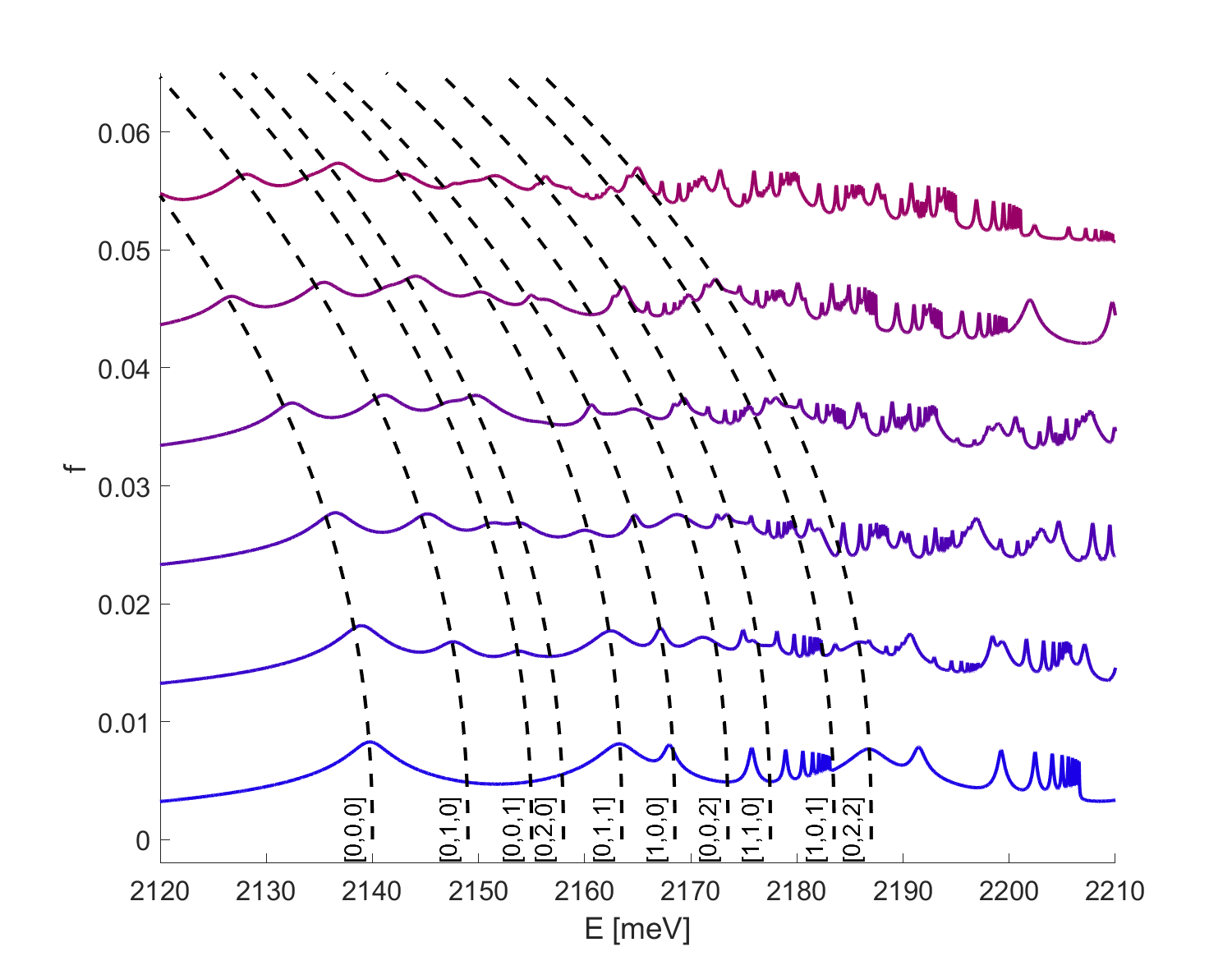}
\caption{The same as Fig. \ref{Fig:1}, shown for a few selected values of $f$. Selected lines are marked with dashed lines and identified.}\label{Fig:2}
\end{figure}

 Fig. \ref{Fig:1} shows the imaginary part of susceptibility calculated from Eq. (\ref{chi2D_Z}) for $L=10$ nm and a range of values of electric field. There is a complicated pattern of absorption lines corresponding to various excitonic states $j=0,1,2...$ and confinement states $N_e=0,1,2...$ $N_h=0,1,2...$. The excitonic number $j$ has the highest impact on linewidth. All resonances experience an energy red-shift proportional to $f^{2}$. One can observe that some states are visible only in some range of values of $f$. In general, lines with higher confinement state numbers are characterized by higher energy and lower amplitude. To identify the particular states, the numbers $[j,N_e,N_h]$ are shown on  Fig.\ref{Fig:2}. The identification of states  becomes very complicated, due to a large number of overlapping peaks, which is also the case in the bulk crystal \cite{Zielinski_2020}, but here  it turned out to be possible to some extend, i.e., for $j=0,1$ to assign quantum numbers to the resonances. The oscillator strength of the basic excitonic states [0,0,0], [1,0,0] etc. decreases with f. For example, $j=1$ exciton ($\sim 2165$ meV, marked [1,0,0]) and $j=2$ exciton ($\sim 2175$ meV) disappear around $f=0.05$ and $f=0.01$, which corresponds to $51$ kV/cm, and $10$ kV/cm. The latter value can be compared with results presented in \cite{Heckotter_2018} and is consistent with them. Overall, the upper limit of considered field values is slightly larger than in available experimental data \cite{Heckotter_2017,Zielinski_2020}.

The first line $[0,0,0]$ is a starting point of several series with increasing $j$, $N_e$, $N_h$. The excitonic states $j$ approach asymptotically a value of $E_g' \approx 2190$ meV, which is the gap energy with additional shift due to the limited thickness $L=10$ nm. The increase of $N_e$ corresponds to the change of energy $\Delta E \sim 8.5$ meV (for example, distance between [0,0,0] and [0,1,0]); the gap between consecutive $N_h$ lines ([0,0,0] and [0,0,1]) is roughly $\Delta E \approx 15$ meV. Again, one can see that some lines are visible only in some range of $f$; for example, [0,1,1] disappears around $f \approx 0.035$. The lines corresponding to high values of $j$, $N_e$, $N_h$ extend beyond the bandgap, creating a very complicated pattern in this region, especially for large values of $f$.

\begin{figure}[ht!]
\centering
\includegraphics[width=1\linewidth]{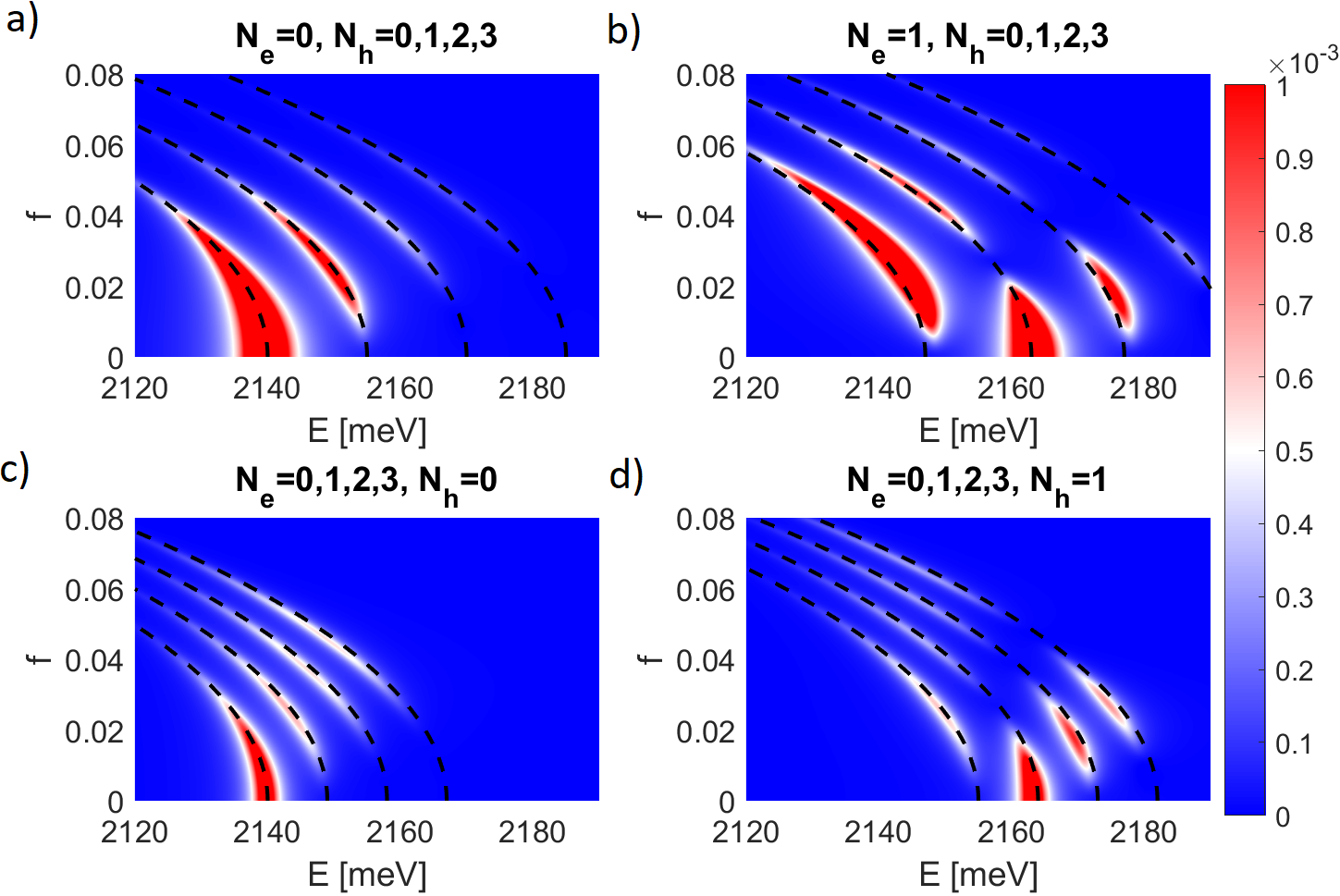}
\caption{Imaginary part of susceptibility calculated for $L=10$ nm, $j=0$ and a) $N_e=0$, b) $N_e=1$, c) $N_h=0$, d) $N_h=1$.}\label{Fig:3}
\end{figure}

To further explore the impact of confinement quantum numbers on the state energy, the susceptibility has been calculated taking into account only $j=0$ excitonic state. The results are shown on the Fig. \ref{Fig:3}.
Four different cases are presented where either $N_e$ or $N_h$ is set to 0 or 1.  Fig. \ref{Fig:3} a) shows the spectrum for $N_e=0$. One can see that the lines corresponding to $N_h=1,2,3,4$ are equally spaced and exhibit the same energy shift with $f$. Higher values of $N_h$ correspond to weaker lines with higher minimal value of $f$ above which the line becomes visible. The only line present at $f=0$ is $N_e=N_h=0$. The spectrum becomes slightly more complex when $N_e=1$, as shown on  Fig. \ref{Fig:3} b); with the exception of $N_h=0$, the lines split into two ranges of $f$ where they have nonzero amplitude. Again, the amplitude decreases with $N_h$ while the energy increases with $N_h$ in a linear manner. Fig. \ref{Fig:3} c) is very similar to Fig. \ref{Fig:3} a), with main difference being smaller energy spacing between $N_e=1,2,3,4$ lines. In the same manner,  Fig. \ref{Fig:3} d) has the same structure as Fig. \ref{Fig:3} b). One can see that only $N_e=N_h$ states are visible for $f=0$. This is also visible on  Fig. \ref{Fig:2}.

\begin{figure}[ht!]
\centering
\includegraphics[width=1\linewidth]{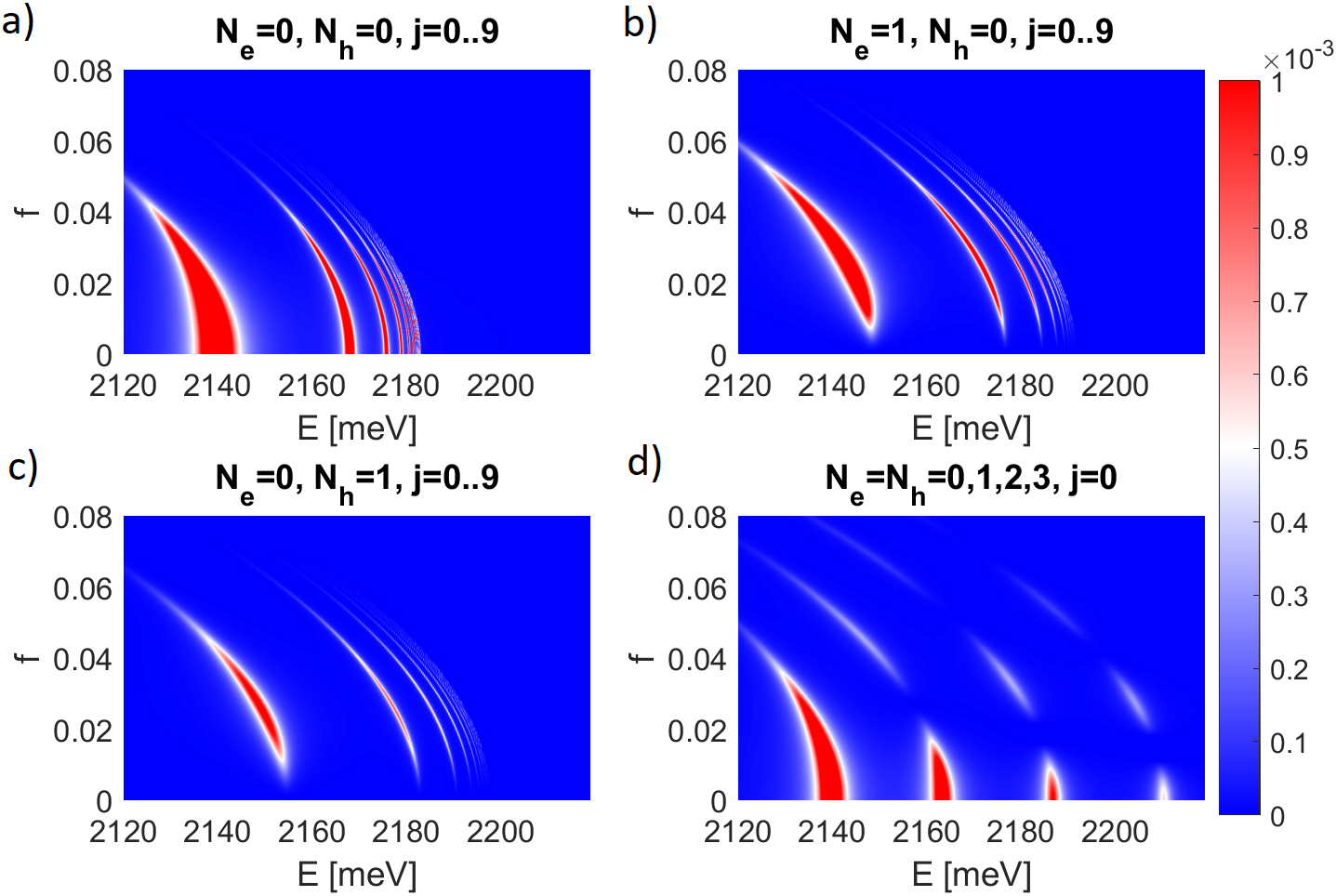}
\caption{Imaginary part of susceptibility calculated for $L=10$ nm, $j=0..9$ and a) $N_e=N_h=0$, b) $N_e=1$, c) $N_h=1$, d) $j=0$,$N_e=N_h=1,2,3,4$.}\label{Fig:4}
\end{figure}

\begin{figure}[ht!]
\centering
\includegraphics[width=1\linewidth]{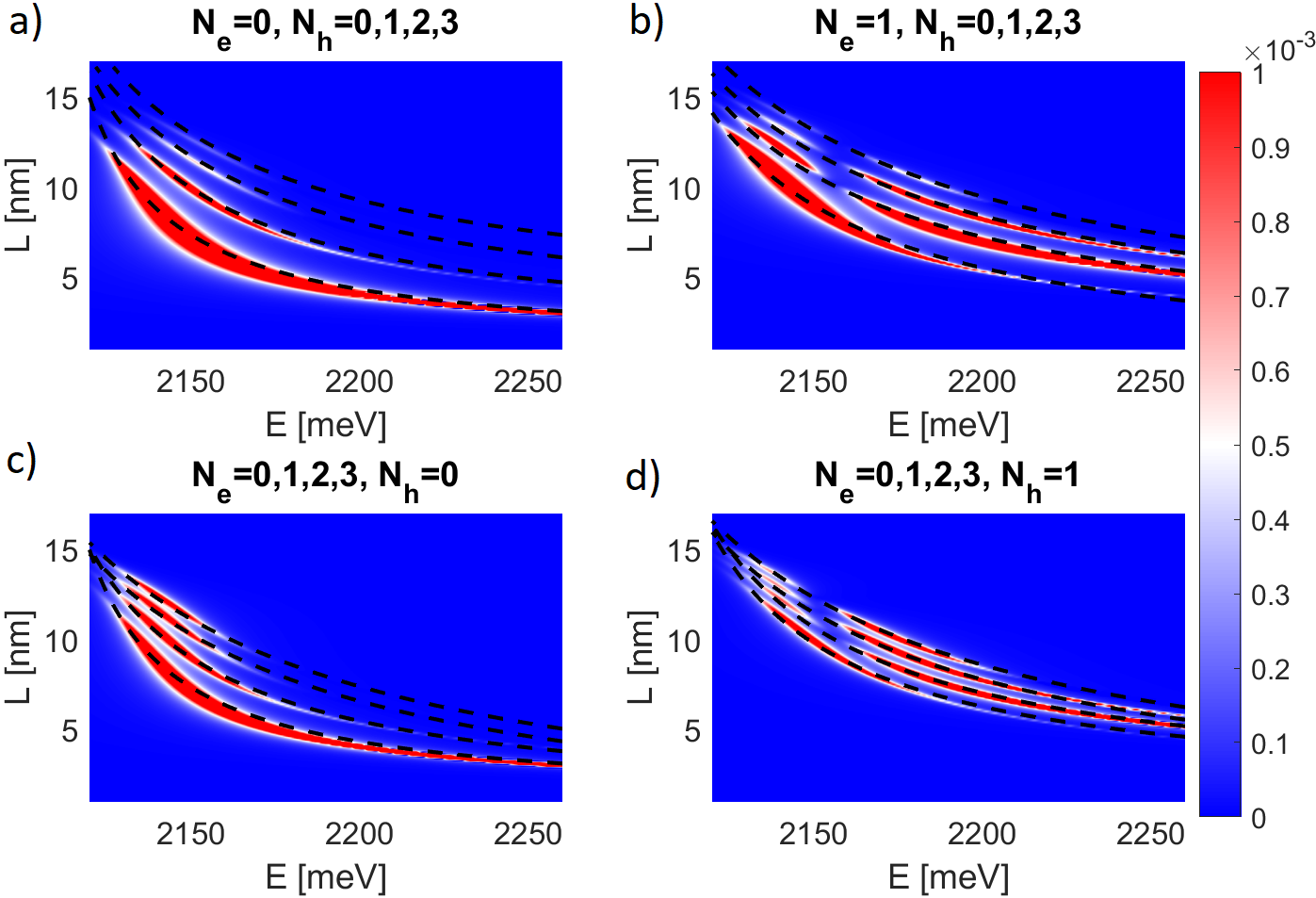}
\caption{Imaginary part of susceptibility calculated for $L=10$ nm, $j=0$ and a) $N_e=0$, b) $N_e=1$, c) $N_h=0$, d) $N_h=1$.}\label{Fig:5}
\end{figure}

The above discussed line series are repeated for every value of excitonic state number $j$. Fig. \ref{Fig:4} a) shows the spectrum calculated for $j=0..9$ and $N_e=N_h=0$. One can see a typical excitonic line series with energy asymptotically approaching some constant value. With increase of either $N_e$ (Fig. \ref{Fig:4} b)) or $N_h$ (Fig. \ref{Fig:4} c)), the whole spectrum is shifted in energy and the range of values of $f$ where the lines are visible moves up. Finally,  Fig. \ref{Fig:4} d) shows the case of various values of $N_e=N_h=1,2,3,4$; one can observe that every consecutive line splits into more separate parts. By observing higher confinement states, we can conclude that for any combination of $N_e$, $N_h$, the line splits into $min(N_e,N_h)+1$ areas where its amplitude is nonzero.
Fig. \ref{Fig:5} shows the dependence on $L$ for the same confinement state combinations as in Fig. \ref{Fig:3}. One can see that in all cases, the energy diverges as $L \rightarrow 0$; however, in contrast to the electric field dependence, the speed of divergence and the exact location of asymptote is different for various values of $N_e$, $N_h$. For example, on  Fig. \ref{Fig:5} a) the line $N_h=0$ approaches infinity as $L \rightarrow 2$ nm, while $N_h=3$ diverges at $L \rightarrow 6$ nm. One can also see that the lines corresponding to higher $N_h$ are present in a narrower range of values of $L$. The spectrum for $N_e=1$ (Fig. \ref{Fig:5} b)) exhibits the same split into two ranges of $L$ as in the case of electric field dependence. The minimum thickness where those lines appear is slightly higher than for $N_e=0$. Again,  Figs. \ref{Fig:5} c) and \ref{Fig:5} d) are analogous to Figs. \ref{Fig:5} a) and \ref{Fig:5} b), but with smaller energy spacing between lines.

\subsection{The \textbf{F} field parallel to the QW layer ($\textbf{F}||x$)}
For the case of electric field parallel to the layer we deal with the effects that are qualitatively similar to those seen in the bulk semiconductor. The main observations are lifting degeneracy of excitonic spectrum due to the external field and appearance of avoided crossings.

\begin{figure}[ht!]
\centering
\includegraphics[width=.9\linewidth]{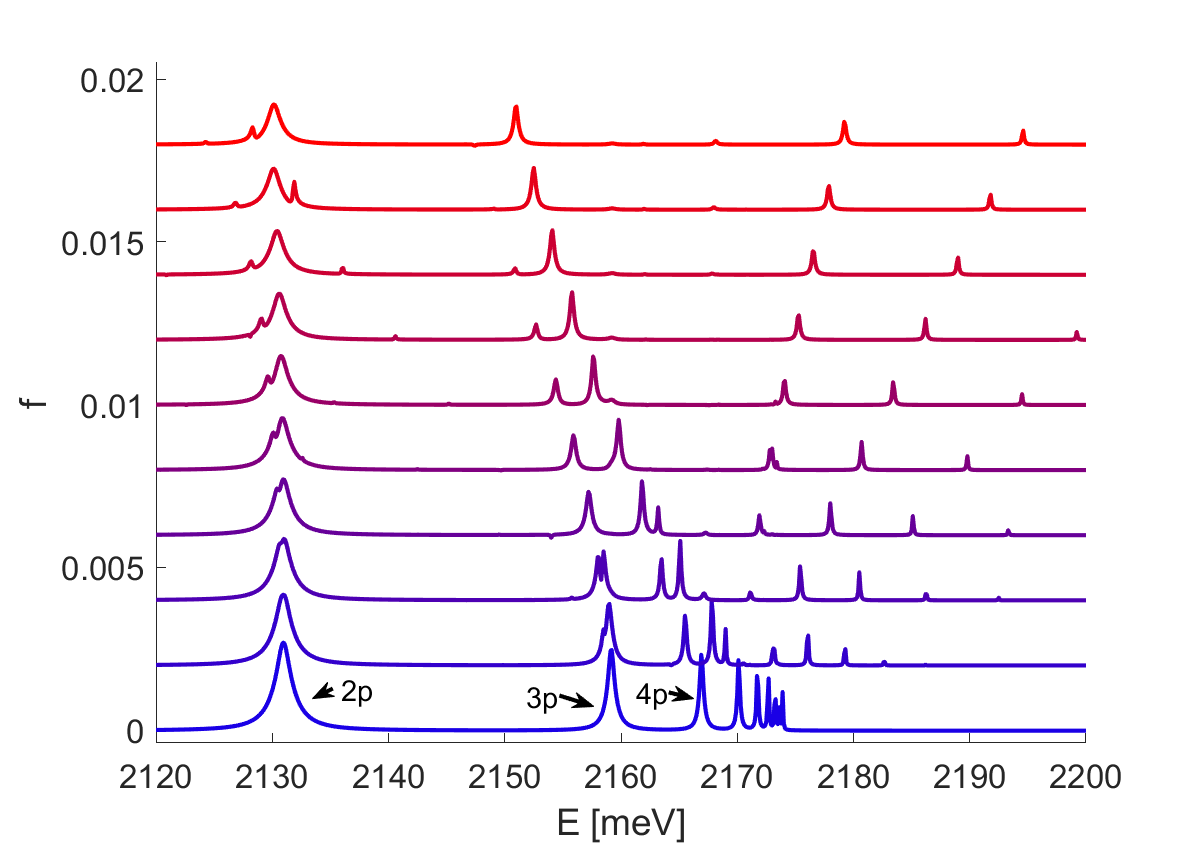}
\caption{Imaginary part of susceptibility as a function of energy and electric field $f$.}\label{Fig:x1}
\end{figure}

\begin{figure}[ht!]
\centering
\includegraphics[width=.95\linewidth]{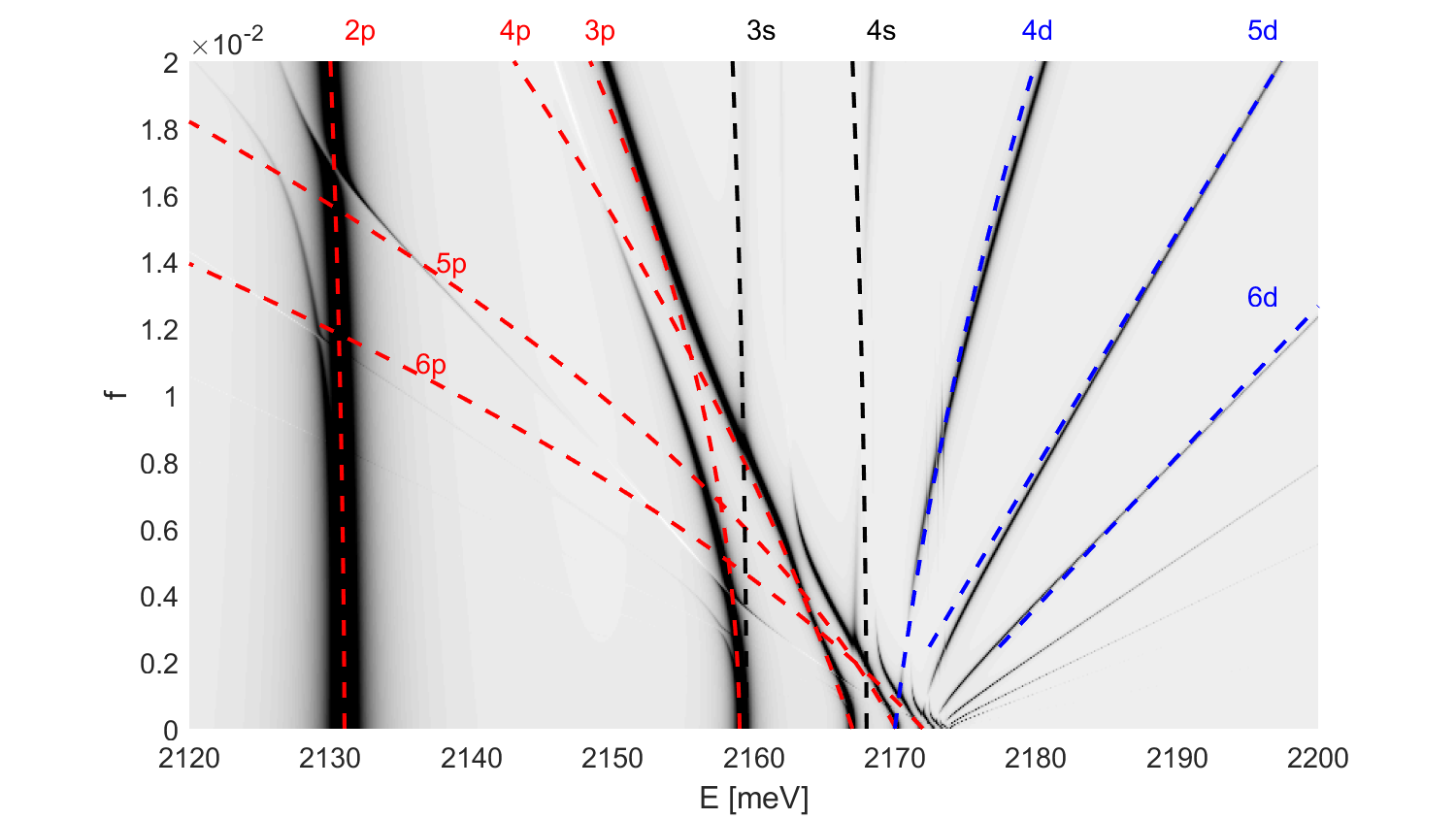}
\caption{Imaginary part of susceptibility (brightness, in log scale) as a function of energy and electric field $f$.}\label{Fig:x2}
\end{figure}

Fig. \ref{Fig:x1} presents the absorption spectrum calculated from Eq. (\ref{chix}) for selected values of electric field $f$ and thickness $L$=20 nm.
At $f=0$, a standard series of p-exciton lines is visible. One can that the exciton energies approach $E \approx 2175$ meV, which is larger than $E_g$ due to the $L$-dependent energy shift. The state $2p$ exhibits very little shift with electric field, but the effect is much stronger for higher states $3p$, $4p$ etc. Due to the fact that the state energy decreases with $f$ and the reduction is faster for upper states, there is a lot of lines overlaps and anticrossing is observed. In the high energy region, multiple small peaks are visible; these maxima correspond to the $d$ excitons, starting from $3d$ levels. Finally, one can observe that the absorption amplitude decreases slowly with electric field.
To better understand the structure of the spectrum, a continuous range of values of $f$ is investigated on  Fig. \ref{Fig:x2}.
The $s$, $p$ and $d$ excitons are marked by black, red and blue lines, accordingly. The $p$ excitons exhibit an approximately quadratic energy shift with $f$; due to the line overlap, only $2p$ and $3p$ excitonic lines are clearly visible in the full range of $f$. One can observe a significant anticrossing of $3p$ and $4p$ lines originating from nondiagonal matrix elements in Eq. (\ref{matrixform}). Interestingly, while the $s$ exciton lines are not highly visible, they also cause anticrossings (for example, intersection of $4s$ and $5p$ lines at $f=4\cdot10^{-3}$). The $d$ exciton lines appear at some minimal value of $f \sim 2\cdot10^{-3}$ and are linearly upshifted with $f$.

\begin{figure}[ht!]
\centering
\includegraphics[width=.8\linewidth]{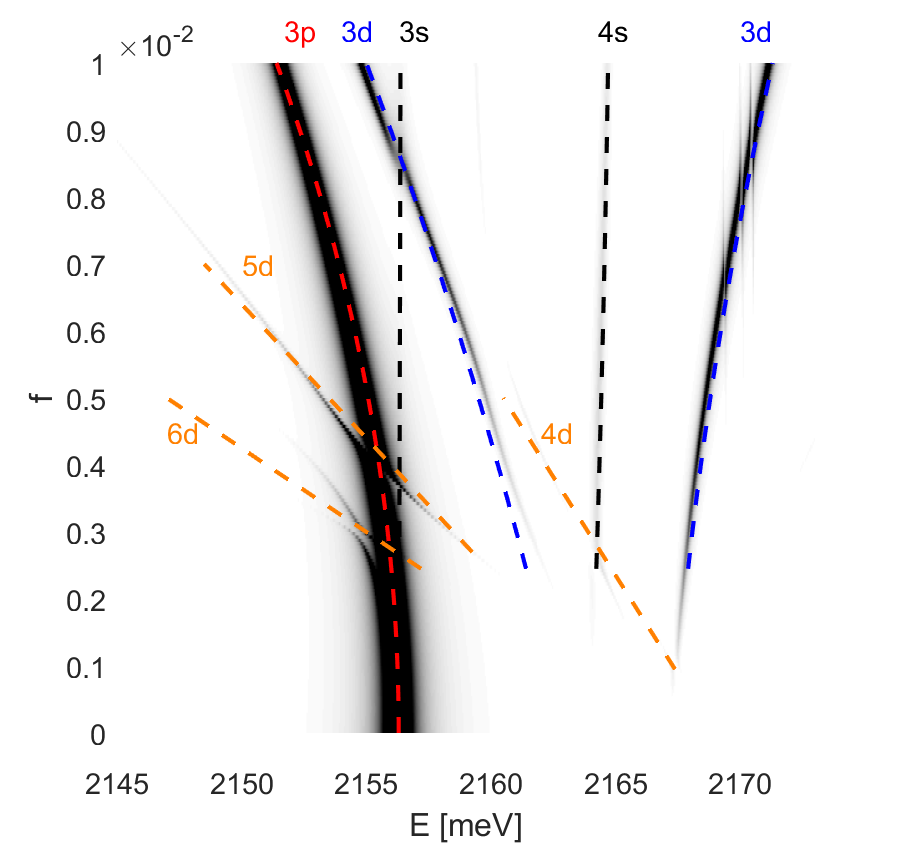}
\caption{Imaginary part of susceptibility (brightness, in log scale) of the\textbf{ n=3} state as a function of energy and electric field $f$.}\label{Fig:x3}
\end{figure}

A more detailed analysis of single excitonic state $j=2$ is shown on  Fig. \ref{Fig:x3}. The overall spectrum structure follows the one presented in \cite{Agekyan1977}; the strongest $3p$ line (red) starts from $E \approx 2157$ meV and exhibits quadratic energy shift. There is a pair of $3d$ lines which originate from a point $E \approx 2165$ meV and split in a linear manner with f. These lines become visible at $f \sim 2\cdot10^-3$; in a high field regime, their amplitude becomes comparable to the $p$ state. Another feature of the spectrum is a $d$ exciton triplet. Those lines are visible mostly in the region of their anticrossing with $p$ and $s$ state. The $3s$ state is also visible for sufficiently strong field; its energy is almost independent of $f$. Note that while for $f=0$ the $s$-exciton has lower energy than $p$ exciton, the situation reverses for high field due to the fact that the $s$ state very weakly affected by the external field in contrast to the $p$ and $d$ states. Such a phenomenon was observed experimentally in \cite{Agekyan1977}.

\begin{figure}[ht!]
\centering
\includegraphics[width=.8\linewidth]{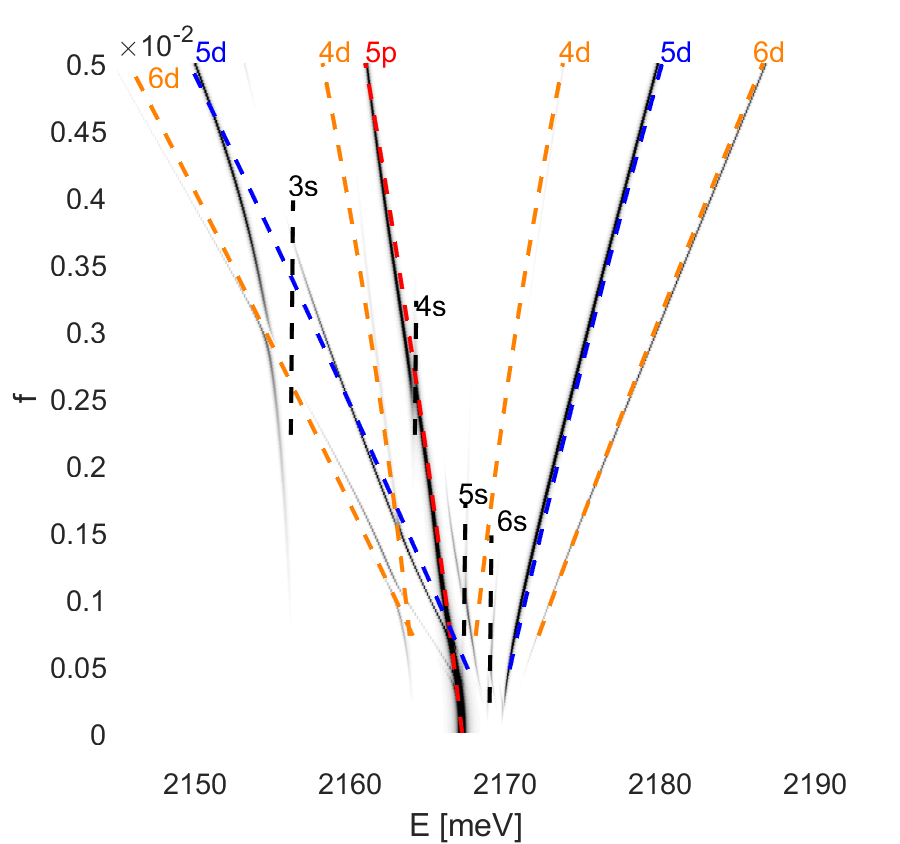}
\caption{Imaginary part of susceptibility (brightness, in log scale) of the \textbf{n=5} state as a function of energy and electric field $f$.}\label{Fig:x4}
\end{figure}

Even more complicated spectrum is obtained for $n=5$ state (Fig. \ref{Fig:x4}). In addition to single $p$, single $s$ and two $d$ states, there are multiple apparent lines that can be attributed to anticrossings with $4d$ and $6d$ excitons. For any given $n$, the states form a structure close to the standard Stark fan \cite{Heckotter_2017} and its width is comparable with experimental results in \cite{Heckotter_2017} for the same electric field value. Again, due to the field-induced downshift, the $5p$ state crosses the lines of $3s$, $4s$ and $5s$ excitons; due to the lower linewidth of $5p$ level, these anticrossings are more apparent.

\begin{figure}[ht!]
\centering
\includegraphics[width=.9\linewidth]{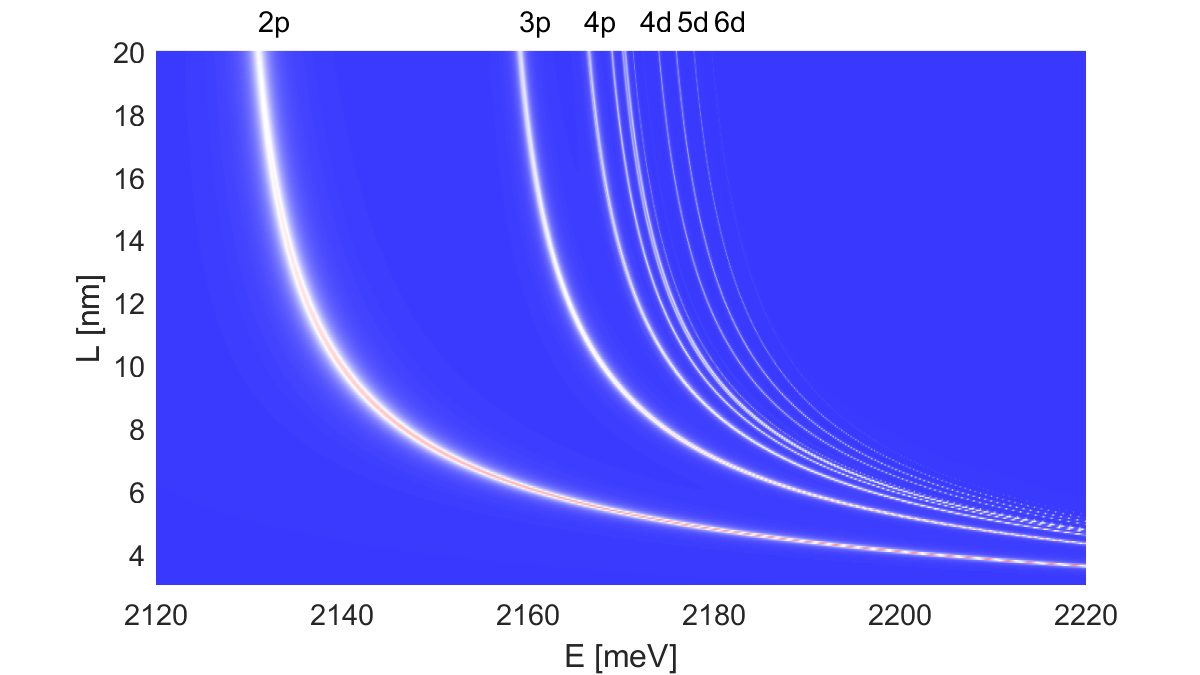}
\caption{Imaginary part of permittivity (color) as a function of energy and well thickness $L$.}\label{Fig:x5}
\end{figure}

Finally,  Fig. \ref{Fig:x5} shows the dependence of exciton energy on the QW thickness $L$, calculated for electric field $f=0.01$. The shift dependence of the lower states is more pronounced  for wider quantum wells. Moreover,  one can see two series of states: the $p$ excitons and the high-energy $d$ excitons. Both types of states exhibit a strong upshift with reduction of L, approaching $E \rightarrow\infty$ as $L \rightarrow 3$ nm. 

\section{Conclusions}
In this paper we have studied the electro-optical properties of Cu$_2$O QWs with Rydberg excitons in two different orientations of the applied  external electric field, for excitation energies below the fundamental gap. For the electric field applied in the z-direction, and in the considered field strengths range, the quadratic Stark red shift of resonance energies prevails, depending on the QW thickness and the total exciton mass. New resonances appear, which are not allowed, for symmetry reasons, when the electric field is absent. We observe even more complicated dependences in the case of the lateral applied field. The resonances can be both red- and blue shifted. We observe a considerable interlevel mixing and splitting caused by differences in energy shifts and various excitonic states due to the interplay between the confinement influence and  the electric field. We believe that tunability of optical properties of QW with RE which is enabled in both electric field configurations makes them suitable for applications as flexible devices in nanotechnology.
\appendix
\section{Quantities $\langle \Psi_{N_eN_h}\rangle_\infty, \langle\Psi_{N_eN_h}\rangle_L$}\label{Appendix A}
We use the definitions (\ref{zetaezetah}), (\ref{definitions}),
and (\ref{definitions_chi}), to calculate the quantities $\langle
\Psi_{NeN_h}\rangle_\infty$, and $\langle \Psi_{NeN_h}\rangle_L$.
We take three combinations: $N_e=0,N_h=0$, $N_e=1, N_h=0$, and
$N_e=0, N_h=1$. Inserting the definitions of the Hermite
polynomials $H_0,H_1$, and performing the respective integrations,
one obtains
\begin{eqnarray}
&&\langle\Psi_{00}\rangle_\infty=\sqrt{\frac{\alpha_e\alpha_h}{p}}\exp\left[-\frac{\alpha_e^2\alpha_h^2(z_{0e}+z_{0h})^2}
{2(\alpha_e^2+\alpha_h^2)}\right],
\end{eqnarray}
\begin{eqnarray}
 &&\langle\Psi_{00}\rangle_L=\sqrt{\frac{\alpha_e\alpha_h}{p}}
 \exp\left[-\frac{\alpha_e^2\alpha_h^2(z_{0e}+z_{0h})^2}
{2(\alpha_e^2+\alpha_h^2)}\right]\\
&&\times\frac{1}{2}\biggl[\hbox{erf}\left(\frac{L\sqrt{p}}{2}+
\frac{q}{\sqrt{p}}\right)
+\hbox{erf}\left(\frac{L\sqrt{p}}{2}-\frac{q}{\sqrt{p}}\right)\biggr],\nonumber
\end{eqnarray}
\begin{eqnarray}
&&\langle\Psi_{10}\rangle_\infty=\frac{\alpha_e}{p}\left(\frac{q}{\sqrt{p}}+z_{0e}\sqrt{p}\right)\sqrt{\frac{\alpha_e\alpha_h}{2}}\\
&&\times \exp\left[-\frac{\alpha_e^2\alpha_h^2(z_{0e}+z_{0h})^2}
{2(\alpha_e^2+\alpha_h^2)}\right],\nonumber\\
&&\langle\Psi_{01}\rangle_\infty=\frac{\alpha_h}{p}\left(\frac{q}{\sqrt{p}}-z_{0h}\sqrt{p}\right)\sqrt{\frac{\alpha_e\alpha_h}{2}}\\
&&\times\exp\left[-\frac{\alpha_e^2\alpha_h^2(z_{0e}+z_{0h})^2}
{2(\alpha_e^2+\alpha_h^2)}\right],\nonumber
\end{eqnarray}
\begin{eqnarray}
&&\langle\Psi_{10}\rangle_L=\nonumber\\
&&=\frac{\alpha_e}{\sqrt{\pi}}\sqrt{\frac{\alpha_e\alpha_h}{2}}
\exp\left[-\frac{\alpha_e^2\alpha_h^2(z_{0e}+z_{0h})^2}
{2(\alpha_e^2+\alpha_h^2)}\right]\\
&&\times\frac{1}{2p}\biggl\{\exp\left[-\left(\frac{L\sqrt{p}}{2}+\frac{q}{\sqrt{p}}\right)^2\right]
-\exp\left[\left(\frac{L\sqrt{p}}{2}-\frac{q}{\sqrt{p}}\right)^2\right]
\nonumber\\
&&+\left(\frac{q}{\sqrt{p}}+z_{0e}\sqrt{p}\right)\sqrt{\pi}\biggl[\hbox{erf}\biggl(\frac{L\sqrt{p}}{2}
+\frac{q}{\sqrt{p}}\biggr)\nonumber\\&&+\hbox{erf}\left(\frac{L\sqrt{p}}{2}-\frac{q}{\sqrt{p}}\right)\biggr]\biggr\}.\nonumber
\end{eqnarray}
and
\begin{eqnarray}
&&\langle\Psi_{01}\rangle_L=\langle e0 \vert h1\rangle\nonumber\\
&&=\frac{\alpha_h}{\sqrt{\pi}}\sqrt{\frac{\alpha_e\alpha_h}{2}}
\exp\left[-\frac{\alpha_e^2\alpha_h^2(z_{0e}+z_{0h})^2}
{2(\alpha_e^2+\alpha_h^2)}\right]\\
&&\times\frac{1}{2p}\biggl\{\exp\left[-\left(\frac{L\sqrt{p}}{2}+\frac{q}{\sqrt{p}}\right)^2\right]
-\exp\left[\left(\frac{L\sqrt{p}}{2}-\frac{q}{\sqrt{p}}\right)^2\right]
\nonumber\\
&&+\left(\frac{q}{\sqrt{p}}-z_{0h}\sqrt{p}\right)\sqrt{\pi}\nonumber\\
&&\times\left[\hbox{erf}\left(\frac{L\sqrt{p}}{2}+
\frac{q}{\sqrt{p}}\right)+\hbox{erf}\left(\frac{L\sqrt{p}}{2}-\frac{q}{\sqrt{p}}\right)\right]\biggr\},\nonumber
\end{eqnarray}
where erf() is the error function \cite{Abramowitz}. The quantities $p$ and $q$ are defined as
\begin{eqnarray}
&&p=\frac{1}{2}\left(\alpha_e^2+\alpha_h^2\right),\nonumber\\
&&q=\frac{1}{2}\left(\alpha_h^2z_{0h}-\alpha_e^2z_{0e}\right).\end{eqnarray}
In all the above expressions, due to Eq. (\ref{alpha}), one has to
put $\alpha_e=\alpha_h$.
\section{Matrix elements for the lateral field}\label{Appendix_B}
For the sake of simplicity we consider only the lowest confinement
state, $N_e=N_h=0$. Using the notation
\begin{eqnarray*}
&&c_{01}=x_1,\;c_{11}=x_2,..,\;c_{J-1,1}=x_J,\;
c_{r-1,1}=x_r,\;r=1,..,J\\
&&c_{00}=x_{J+1},\;c_{10}=x_{J+2},..,\;c_{r-1,0}=x_r,\quad
r=J+1,..,\;2J,\\
&&c_{02}=x_{2J+1},\;c_{12}=x_{2J+2}\quad c_{r-1,2}=x_r,\quad
r=2J+1,..,\;3J.\\
&&a_{ij}=\kappa_{i-1,1}^2\delta_{ij},\quad i=1,.., J,\quad j=1,.., J\\
&&a_{ij}=V^{01}_{i-J-1,j-1},\quad i=J+1,.., 2J,\\
&&a_{ij}=V^{21}_{i-2J-1,j-1},\quad i=2J+1,.., 3J.
\end{eqnarray*}
we put equations (\ref{eqslateral})  into a matrix form
\begin{eqnarray*}
&&\underline{\underline{A}}\textbf{X}=\textbf{b},\\
&&\textbf{X}=(x_1,x_2,\cdots x_{3J}),\\
&&\textbf{b}=(b_1,b_2,\cdots b_{3J}), \end{eqnarray*} where the
matrix elements $\underline{\underline{A}}$ are defined as
follows
\begin{eqnarray*}
&&i,j=1,\cdots J,\quad a_{ij}=\kappa_{i-1,1}^2\delta_{ij},\\
&&\kappa_{i-1,1}^2=\frac{1}{R^*}\left(E_g-\hbar\omega-i{\mit\Gamma}+\epsilon_iR^*+W_{e0}+W_{h0}\right),\\
&&\epsilon_i=-4\lambda_i^2,\\
&&i=1,\cdots J,\quad j=J+1,\cdots 2J, \quad a_{ij}=V^{10}_{i-1,j-J-1},\\
&&i=1, \cdots J,\quad j=2J+1,\cdots 3J, \quad a_{ij}=V^{12}_{i-1,j-2J-1}.
\end{eqnarray*}
\begin{eqnarray*}
&&i=J+1,\cdots 2J, \quad j=1,\cdots J,\quad a_{ij}=2V^{01}_{i-J-1,j-1},\\
&&j=J+1,\cdots 2J, \quad a_{ij}=\kappa_{j-J-1,0}^2\delta_{ij},\\
&&i'=j-J-1,\\
&&\kappa_{i',0}^2=\frac{1}{R^*}\left(E_g-\hbar\omega-i{\mit\Gamma}+\epsilon_{i'}R^*+W_{e0}+W_{h0}\right),\\
&&\epsilon_{i'}=-\frac{4}{(2i'-1)^2}R^*, \quad i'=1,\cdots, J,\\
&&j=2J+1,\cdots 3J,\quad a_{ij}=0,
\end{eqnarray*}
\begin{eqnarray*}
&&i=2J+1,\cdots 3J,\quad j=1,\cdots J,\quad a_{ij}=2V^{21}_{i-2J-1,j-1},\\
&&j=J+1,\cdots, 2J,\quad a_{ij}=0,\\
&&j=2J+1,\cdots 3J, \quad a_{ij}=\kappa_{j-2J-1,2}^2\delta_{ij},\\
&&i''=j-2J-1,\\
&&\kappa_{i'',0}^2=\frac{1}{R^*}\left(E_g-\hbar\omega-i{\mit\Gamma}+\epsilon_{i''}R^*+W_{e0}+W_{h0}\right),\\
&&\epsilon_{i''}=-\frac{4}{(2i''+3)^2}R^*, \quad i''=1,\cdots, J,\\
\end{eqnarray*}
The coefficients $b_1,\ldots,b_{3J}$ are defined as
\begin{eqnarray*}
&&i=1,\cdots, J,\\
&&b_i=\sqrt{\frac{i(i+1)}{(i+\frac{1}{2})^5}}(1+2\rho_0\lambda_{i})^{-4}
F\left(1-i,4;3;\frac{1}{s}\right),\nonumber\\
&&\lambda_i=\frac{1}{2i+1},\\
&&s=\frac{1+2\rho_0\lambda_i}{4\rho_0\lambda_i},\\
\end{eqnarray*}
and elements $V_{ij}$ are given by following formulas
\begin{eqnarray}
&&V_{ij}^{10}=\frac{1}{2}f\frac{4^3\lambda_{i1}}{[(2i+3)(2j+1)]^{3/2}}\left[\frac{i!}{(i+2)!}\right]^{1/2}\\
&&\times\sum\limits_{r=0}^i\sum\limits_{s=0}^j
(-1)^{r+s}{i+2\choose i-r}{j\choose
s} \nonumber \\ 
&&\times \frac{(4\lambda_{i1})^r(4\lambda_{j0})^s}{r!s!}\frac{(r+s+3)!}{[2(\lambda_{i1}+\lambda_{j0})]^{r+s+4}},\nonumber\end{eqnarray}
\begin{eqnarray}
&&V_{ij}^{01}=\frac{1}{2}f\,\frac{4^3\lambda_{j1}}{\left[(2i+1)(2j+3)\right]^{3/2}}\left[\frac{j!}{(j+2)!}\right]^{1/2}\\
&&\times\,\sum\limits_{r=0}^i{i\choose
r}\sum\limits_{s=0}^j(-1)^{r+s}{j+2\choose
j-s}\frac{(4\lambda_{i0})^r(4\lambda_{j1})^s(r+s+3)!}{r!s![2(\lambda_{i0}+\lambda_{j1})]^{r+s+4}},\nonumber
\end{eqnarray}
\begin{eqnarray}
&&V_{ij}^{12}=\frac{512
f}{[(2i+3)(2j+5)]^{3/2}}\left[\frac{i!j!}{(i+2)!(j+4)!}\right]^{1/2}\frac{1}{(2i+3)}\nonumber\\
&&\times\left[\frac{1}{(2j+5)}\right]^2\sum\limits_{r=0}^i\sum\limits_{s=0}^j(-1)^{r+s}{i+2\choose
i-r}{j+4\choose
j-s}\nonumber\\
&&\times
\frac{(4\lambda_{i1})^r(4\lambda_{j2})^s}{r!s!}\frac{(r+s+5)!}{[2(\lambda_{i1}+\lambda_{j2})]^{r+s+6}},\end{eqnarray}
\begin{eqnarray} &&V_{ij}^{21}=\frac{512
f}{[(2i+5)(2j+3)]^{3/2}}\left[\frac{i!j!}{(i+4)!(j+2)!}\right]^{1/2}\frac{1}{(2i+5)^2}\nonumber\\
&&\times\frac{1}{(2j+3)}\sum\limits_{r=0}^i\sum\limits_{s=0}^j(-1)^{r+s}{i+4\choose
i-r}{j+2\choose
j-s}\nonumber\\
&&\times\frac{(4\lambda_{i2})^r}{r!}\frac{(4\lambda_{j1})^s}{s!}\frac{(r+s+5)!}{[2(\lambda_{i2}+\lambda_{j1})]^{r+s+6}}.
\end{eqnarray}

\end{document}